\documentclass[sigconf]{acmart}
\setcopyright{none}
\renewcommand\footnotetextcopyrightpermission[1]{}

\usepackage{booktabs}

\begin{document}

\title{Claimed or Attested?\\
       A Commit-Signature Dataset and Identity Trust Tiers
       across the World of Code}

\author{Audris Mockus}
\affiliation{%
  \institution{University of Tennessee, Knoxville}
  \city{Knoxville}\state{TN}\country{USA}}
\email{audris@utk.edu}

\begin{abstract}
An author string in a git commit is free text the committer typed, so identity
resolution over a global commit corpus rests on a claim that nothing in the
commit verifies. A cryptographically signed commit is different: it binds the
commit to a key the committer controls, and when that key ties back to a
real-world identity the git identity becomes attested rather than merely claimed.
We release the first commit-signature axis for the World of Code (WoC), extracted
for the \texttt{V2604} collection. The signature travels in the commit object's
\texttt{gpgsig} header and is already carried, unparsed, in the commit-message
field of the WoC commit tables, so the axis is a scan over existing tables rather
than a re-read of the object database. Over the \texttt{V2604} corpus of
$5{,}866{,}595{,}698$ commits, $17.59\%$ carry a signature (PGP dominant at
$98.96\%$, with a growing minority of SSH and X.509/sigstore signatures), or
$1{,}031{,}721{,}316$ signed commits. We release the per-commit signature map \texttt{c2sigFull}, a
key-to-author graph gated so that shared organization and continuous-integration
keys are separated from person keys, and \texttt{A2trust}, a per-identity
attestation tier (unsigned, signed, real-world-bound, cross-corpus attested) that
extends the published \texttt{A2cls} identity-class dataset. The signature axis is
a precision anchor, not a coverage layer: signed commits skew toward
security-conscious developers, a population that overlaps the scholarly authors a
bibliography join targets. We use the person keys to build a cryptographically
grounded alias gold that calibrates the heuristic WoC alias map independently of
hand-labeled pairs, and to attach an attestation provenance to science-to-software
identity links. All artifacts are released as a self-contained, independently
hosted replication package keyed to the WoC \texttt{V2604} collection.
\end{abstract}


\maketitle

\section{Introduction}
\label{sec:intro}
Every study that mines a global commit corpus has to decide when two author
strings denote the same person and when one string denotes two. The World of Code
alias map, like other disambiguation systems, resolves this from co-occurrence of
names, emails, and GitHub logins~\cite{amreen2019alfaa}. Those signals are useful,
but they share a blind spot: the author field of a commit is free text the
committer chose, so every merge and every split the map produces rests on a claim
that nothing in the commit itself corroborates. Vanity strings make the gap
concrete. Unrelated people commit as \texttt{root}, and well-known names and
addresses are reused by impostors; the WoC bad-identity stoplist literally
enumerates such strings. A name in a commit is a claim, not an attestation.

Scholarly authorship sits at the opposite corner of the same problem. A paper
carries the author's real name because reputation is the point of publishing, so
impersonation is rare and the dominant error is homonymy: many people named
J.~Smith, or one person whose transliterated name is split across spellings. When
a study links software authors to paper authors, it joins a low-impersonation,
high-homonym universe to a high-impersonation, high-homonym one, and the join
inherits the weaker guarantee.

A cryptographically signed commit changes the epistemics on the software side. Git
supports signing a commit with a PGP key, an SSH key, or an X.509 certificate; the
signature binds the commit content to a key the committer controls. If the key
ties to a real-world identity, through a PGP user-id email, an SSH key registered
on a GitHub account, or an X.509 or OIDC subject, then the git identity used on
that commit is attested at the same level the paper side enjoys, and the
bibliography join becomes defensible rather than heuristic. This is the
opportunity the dataset captures.

\begin{figure}[t]
\centering
\includegraphics[width=\linewidth]{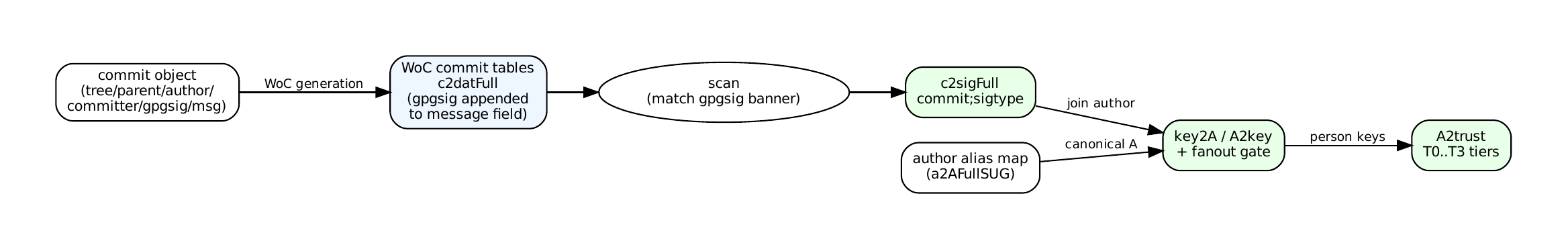}
\caption{The signature already lives in the commit tables. Git writes
\texttt{gpgsig} between the \texttt{committer} line and the message; the WoC table
generator appends that block to the message field, so a scan over the existing
commit tables yields the signature axis. Keys are then joined to authors and gated
so that shared organization and CI keys do not act as person anchors.}
\label{fig:attestation}
\end{figure}

We release the first commit-signature axis for World of Code~\cite{ma2019woc,
ma2021world}. The construction rests on an observation about how WoC already
stores commits: the \texttt{gpgsig} header, which git places between the
\texttt{committer} line and the message, survives into the message field of the
WoC commit tables because the table generator appends every post-\texttt{committer}
header line to the message. The signature is therefore already on disk, and
extracting it is a scan over the existing \texttt{V2604} commit tables rather than
a large-RAM pass over the object database (Figure~\ref{fig:attestation}). We
document the construction and results as an experiment log (Exps.~S1--S5). Our
contributions are the released artifacts and the findings about them:
\begin{itemize}
\item \textbf{A per-commit signature map} \texttt{c2sigFull}, labeling each signed
  commit with its signature family, extracted for the whole \texttt{V2604} corpus
  from the existing commit tables with no object-database pass (Exp.~S1), together
  with the prevalence of commit signing overall and by identity class (the adoption
  curve over author-time is a planned extension).
\item \textbf{A key-to-identity graph} \texttt{key2A}/\texttt{A2key} with a
  key-fanout gate that separates person keys from shared organization and
  continuous-integration keys, the signing analogue of the name-spread gate that
  identity disambiguation already relies on (Exp.~S2).
\item \textbf{A cryptographically grounded alias gold} built from author strings
  co-signed by the same person key, used to calibrate the heuristic WoC alias map
  independently of hand-labeled pairs (Exp.~S3).
\item \textbf{An identity trust-tier map} \texttt{A2trust} (unsigned, signed,
  real-world-bound, cross-corpus attested) that extends the published
  \texttt{A2cls} identity-class dataset with a principled confidence axis, and a
  signature-based test for impersonation on vanity strings (Exp.~S4).
\item \textbf{A bibliography attestation bridge} that upgrades science-to-software
  identity links from name-match heuristics to cryptographic anchors where a
  signing key resolves to a scholarly author, with a sampled verification pass
  that bounds the residual forgery rate (Exp.~S5).
\end{itemize}

The dataset is a drop-in join key for any WoC-scale study: per-commit signatures
join the commit tables by SHA, and per-identity tiers join the author summary. It
does not replace the heuristic alias map; it gives that map an independent,
cryptographic check where signing coverage allows, and it marks exactly which
identities carry an attestation a downstream analyst can rely on.

\section{Related Work}
\label{sec:related}
The corpus we extend is World of Code~\cite{ma2019woc,ma2021world}, a mirror of
public version-control data organized as commit, tree, blob, and relation maps.
Identity resolution over it is the ALFAA line of work~\cite{amreen2019alfaa},
which disambiguates authors from behavioral and textual co-occurrence of names,
emails, and logins, and which we take as the heuristic map this dataset
calibrates. The vanity-string and impersonation problem those methods contend with
is the motivation for an attested axis: a claim-based resolver cannot, on its own,
tell a real owner from an impostor reusing the same string.

Commit signing itself has grown from a niche practice to a platform-supported
default. Git added SSH-key signing in 2.34, GitHub verifies and displays signature
status, and sigstore's \texttt{gitsign}~\cite{sigstore} introduced keyless signing
backed by short-lived Fulcio certificates whose subject is a verified OIDC
identity. These developments make signatures both more common and more directly
tied to a real-world identity than classical PGP web-of-trust, which is what makes
a signature axis worth extracting now. We are, to our knowledge, the first to
extract and release commit signatures at the scale of a global corpus and to model
them as an identity-trust layer rather than a per-repository security check.

The dataset is a sibling of two other World of Code data
showcases that also confront corpus-level identity and completeness questions: the
history-rewrite provenance dataset~\cite{historyrewriteshowcase}, which separates
never-ingested commits from upstream-rewritten ones, and the deforking
map~\cite{forks20,deforkingshowcase}, which resolves which repositories are the
same project. Where those resolve provenance of commits and of repositories, this
one resolves provenance of the \emph{identity} attached to a commit. It also
connects to cross-corpus author linking~\cite{bock2025dealiasing}, supplying a cryptographic
provenance for the science-to-software identity edges that work builds by name and
ORCID matching.


\section{Construction}
\label{sec:construction}
A signed git commit places a \texttt{gpgsig} header after the \texttt{committer}
line and before the blank line that separates the header from the message; the
signature's continuation lines are space-prefixed, so the armored block, including
its own internal blank line encoded as a single space, does not break the header.
Three signature families occur. PGP signatures (\texttt{-----BEGIN PGP
SIGNATURE-----}) carry an issuer key-id subpacket that is readable without the
public key. SSH signatures (the \texttt{SSHSIG} format, \texttt{-----BEGIN SSH
SIGNATURE-----}) embed the signer's public key, so a fingerprint is derivable with
no network access. X.509 and sigstore signatures (\texttt{-----BEGIN SIGNED
MESSAGE-----}) carry a subject email or, for \texttt{gitsign}, a short-lived
Fulcio certificate whose subject-alternative-name is a verified OIDC identity.

The extraction rests on how World of Code already stores commits. The commit-table
generator decodes each commit object and splits it on the first blank line into a
header block and a message; while parsing the header it appends every line after
the \texttt{committer} line to the message list. For a signed commit that trailing
block is exactly the \texttt{gpgsig} armored signature, so the signature is carried
verbatim, if unparsed, in the message field of the commit tables
(\texttt{c2datFull}, one row per collected commit). Recovering the signature axis
is therefore a scan over the existing \texttt{V2604} commit tables, matching the
canonical header marker \texttt{gpgsig -----BEGIN} and its family banner, and not a
re-read or re-decompression of the object database. The extractor runs as a
sharded batch job over the 128 commit-table shards on commodity nodes at 8\,GB, with
no large-memory or object-database exception. The steady-state path is to capture
the same header during commit-table generation so the axis rides ingestion; for
this release we extract it from the current tables.

Each signed commit contributes one row \texttt{commit;sigtype} to
\texttt{c2sigFull}, sorted by commit SHA and sharded to match the commit tables.
Because the signed commit object already carries its own \texttt{author} line, no
commit-to-author join is needed to attribute a signature; the author string is in
the same row of the source table, and the canonical identity follows from the
existing author alias map.

\section{Signature Inventory and Prevalence (Exp.~S1)}
\label{sec:s1}
The headline data contribution is the prevalence of commit signing across the
corpus. Scanning all $128$ commit-table shards, $1{,}031{,}721{,}316$ of the
$5{,}866{,}595{,}698$ commits in \texttt{V2604} carry a signature, a corpus rate of
$17.59\%$. The signed share is stable across SHA-space shards (three shards drawn
from opposite ends of the hash range give $17.58\%$, $17.53\%$, and $17.65\%$), and
the strict header marker and the loose family banner agree to within $0.01$
percentage points, confirming that the matches are genuine header signatures rather
than signatures quoted inside a commit message. The three families are separated in
Table~\ref{tab:sigfamily}: PGP dominates at $98.96\%$ of signed commits, SSH follows
at $1.02\%$, and X.509/sigstore is a small but growing share ($0.02\%$) concentrated
in recent history, consistent with SSH signing arriving in git 2.34 and keyless
signing arriving later still.

\begin{table}[t]
\centering
\caption{Signature family split across all $1{,}031{,}721{,}316$ signed commits in
\texttt{V2604} (exact, full corpus). The signed rate is $17.59\%$ of
$5{,}866{,}595{,}698$ commits.}
\label{tab:sigfamily}
\small
\begin{tabular}{@{}lrr@{}}
\toprule
family & signed commits & share of signed \\
\midrule
PGP (\texttt{-----BEGIN PGP SIGNATURE})    & $1{,}021{,}000{,}040$ & $98.96\%$ \\
SSH (\texttt{SSHSIG})                       & $10{,}496{,}637$      & $1.02\%$  \\
X.509 / sigstore (\texttt{SIGNED MESSAGE})  & $224{,}639$           & $0.02\%$  \\
\midrule
total signed                                & $1{,}031{,}721{,}316$ & $100\%$   \\
\bottomrule
\end{tabular}
\end{table}

Prevalence alone is a completeness fact; the more useful question is \emph{who}
signs. Mapping the distinct signing ids on a representative shard through the
published \texttt{A2cls} identity-class dataset (Table~\ref{tab:sigclass}), signers
are heavily concentrated in the clean population: $99.52\%$ fall in the good
(developer) class against a corpus baseline of $94.37\%$, and every low-quality
class is under-represented, most sharply the machine-\emph{local} ids at $0.005\%$
of signers versus $2.40\%$ of the corpus, a roughly $500\times$ gap. The whole
non-good tail is $0.37\%$ of signers against $5.11\%$ of ids at large, so the
signing population is about $14\times$ cleaner than the corpus. This is the
selection effect the paper leans on: a signature does not reach the hard-to-resolve
generic and machine-local ids, it marks a high-precision core.

\begin{table}[t]
\centering
\caption{Identity-class composition of the distinct commit signers on a
representative SHA shard, against the full-corpus \texttt{A2cls} baseline. Signers
concentrate in the good/developer class; the low-quality tail is heavily
under-represented.}
\label{tab:sigclass}
\small
\begin{tabular}{@{}lrr@{}}
\toprule
class & share of signers & corpus baseline \\
\midrule
good / developer & $99.52\%$ & $94.37\%$ \\
bad-by-attribute & $0.36\%$  & $2.48\%$  \\
bot              & $0.11\%$  & $0.52\%$  \\
partial          & $0.007\%$ & $0.23\%$  \\
local            & $0.005\%$ & $2.40\%$  \\
\bottomrule
\end{tabular}
\end{table}

The distinct-signer counts behind Table~\ref{tab:sigclass} are drawn from one
representative shard and are size-biased in absolute terms, though stable as class
fractions; the exact global distinct-signer count follows from the full 128-shard
key-graph build (Exp.~S2). The adoption curve over commit author-time and the
signed-rate over the science-repository subset from the cross-corpus linkage are
deferred to that build.

The selection bias is stated up front and carried through every later claim: signed
commits skew toward security-conscious and high-profile developers, which
is close to the population a bibliography join targets but is not representative of
the corpus as a whole. The signature axis is a precision anchor and a validation
layer, never a coverage layer.

\section{Key-to-Identity Graph and the Fanout Gate (Exp.~S2)}
\label{sec:s2}
Joining \texttt{c2sigFull} to the commit identity tables collects, per key handle,
the set of author strings, committer strings, GitHub logins, and projects the key
signs under. Recovering the key handle needs the signature packet, not just its
banner: \texttt{pgpissuer.pl} parses the OpenPGP signature packet to its issuer
subpacket, preferring the 20-byte issuer fingerprint (subpacket 33) and falling back
to the 8-byte issuer key-id (subpacket 16), normalizing both to the low 16 hex
digits (\texttt{keyid16}) that the two forms share. Validated against
\texttt{gpg}'s own listing on a sample, the parser resolves an issuer for $99.92\%$
of parsed PGP signatures.

A key that signs commits authored by many distinct people is not a person anchor:
the GitHub web-flow signing key, which signs web-edit and merge commits on behalf of
every user, is the canonical example, and shared deploy keys and continuous-integration
signing identities behave the same way. The concentration is heavy. Across the full
corpus, $39{,}077{,}350$ signed authors associate with $586{,}011$ keys through
$44{,}871{,}077$ author--key pairs; the single web-flow key \texttt{4AEE18F83AFDEB23}
binds to $30{,}440{,}944$ distinct authors and one further platform key to
$12{,}592{,}731$, so two shared keys carry $95.9\%$ of all author--key associations. We
apply a key-fanout gate on the number of distinct authors per key, the direct analogue of
the name-spread gate that author disambiguation uses to tell anchors from bridges, and
label each key as a person key or a shared organization/CI key. A threshold of $50$
distinct authors drops the $2{,}651$ shared keys ($0.45\%$ of keys, yet $97.6\%$ of
associations) and leaves $583{,}360$ person keys, $72.20\%$ ($423{,}095$) of them bound to
a single author. The surviving person keys give, per key, a cluster of author strings that
is high-trust same-identity evidence independent of the name and email heuristics:
$156{,}397$ person keys (fanout $2$--$20$) co-sign under $549{,}388$ canonical author ids,
i.e.\ $392{,}991$ candidate same-person merges, author strings the alias map holds apart
yet one keyholder demonstrably controls. The outputs are \texttt{key2AFull} (key to
authors, with the person/shared label), its inverse \texttt{A2keyFull}, and
\texttt{key2fanoutFull} ($586{,}011$ keys); the SSH-pubkey and X.509-subject handle
families are the remaining extension.

\section{Signature-Attested Alias Gold and Map Calibration (Exp.~S3)}
\label{sec:s3}
From the person keys we build a cryptographic alias gold: pairs of distinct author
strings co-signed by the same person key are the same individual, with a guarantee
that does not depend on any name or email similarity. This set is larger and
independently grounded compared with the hand-labeled alias pairs used to validate
the current map. We measure the heuristic WoC alias map against it as precision (of
the map's merges, the fraction a shared person key corroborates) and recall (of the
key-co-signed pairs, the fraction the map already merges), and we triage the
disagreements. Pairs co-signed by one key but left unmerged are recall gaps and
candidate new edges; pairs merged by the map whose commits carry different keys are
either key rotation, one person holding several keys over time, or genuine
over-merges, which the key evidence helps separate.

The person keys yield $160{,}265$ signing keys of fan-out 2 to 50 that co-sign
$669{,}147$ distinct raw author strings; every string resolves in the production map,
so there is no coverage loss, and the co-signed set is larger than the $469$k
hand-labeled ALFAA pairs while needing no manual labeling. Reading the map against
this gold as recall (Table~\ref{tab:calib}), the fraction of key-co-signed strings the
map already merges falls monotonically with key fan-out, from $0.630$ at fan-out~2 to
$0.014$ at fan-out 21 to 50. That decline is the team-key confound made visible: a
high-fan-out key is shared (CI, deploy, or team), so its co-signed strings are
different people the map correctly keeps apart, and only the low-fan-out tier is
trustworthy evidence of a single keyholder.

\begin{table}[t]
\centering
\caption{Map recall against the signature-attested gold, by key fan-out. Pair-recall
is the fraction of key-co-signed raw author-string pairs the production map already
merges. The monotone decline tracks shared team/CI keys, not a map defect.}
\label{tab:calib}
\small
\begin{tabular}{@{}lrrr@{}}
\toprule
fan-out & keys & map unifies & pair-recall \\
\midrule
2      & $86{,}929$ & $63.0\%$ & $0.630$ \\
3      & $28{,}176$ & $33.6\%$ & $0.460$ \\
4--5   & $20{,}316$ & $12.8\%$ & $0.281$ \\
6--10  & $14{,}262$ & $1.8\%$  & $0.117$ \\
11--20 & $6{,}714$  & $0.1\%$  & $0.041$ \\
21--50 & $3{,}868$  & $0.0\%$  & $0.014$ \\
\bottomrule
\end{tabular}
\end{table}

Signatures alone are not a clean gold, so we triage the fan-out-2 disagreements. Of
the $32{,}192$ fan-out-2 candidate merges the map left apart, only $9{,}036$
($28.1\%$) carry a corroborating same-person signal, that is a shared email
local-part, a shared non-generic domain, or a shared name token; these are genuine
recall-repair edges the map missed. The other $23{,}156$ ($71.9\%$) are disjoint
identities, two people sharing one key, which the map correctly did not merge.
Corroboration is therefore required: the raw fan-out-2 miss rate overstates the recall
gap, and the corroborated $9{,}036$ edges (\texttt{recall\_repair\_confirmed.tsv}) are
the high-precision candidates to feed back into the map.

\section{Trust Tiers and Impersonation (Exp.~S4)}
\label{sec:s4}
We assign each canonical identity an attestation tier and release it as
\texttt{A2trust}, an evidence field that extends the published \texttt{A2cls}
identity-class dataset. Tier~T0 is unsigned, the current default, a claim only.
Tier~T1 is signed by some key with commits consistent with a single keyholder over
that string. Tier~T2 adds a real-world binding: a PGP user-id email, an SSH key
registered on a GitHub account, or an X.509/OIDC subject. Tier~T3 is
cross-corpus attested, where the key or identity also resolves to a paper author or
ORCID (Exp.~S5). The tier gives every downstream consumer a principled confidence
axis on identity, orthogonal to the developer/bot/bad/local class already published.

The same person keys give a positive test for the impersonation that vanity strings
invite. For a high-profile author string, the real owner's signed commits establish
a dominant key; commits bearing the same string but unsigned, or signed by an
unrelated key, are suspect. Where signing coverage allows, this replaces an entry
on the hardcoded bad-identity stoplist with a signature-derived test, and it yields
a per-string key-dispersion measure that feeds attestation-based bad-identity
detection back into the disambiguation pipeline.

Across the $62{,}579{,}994$ canonical authors the tiers split as
Table~\ref{tab:tiers}: $49.0\%$ hold no signed commit (T0), $50.3\%$ touched a
platform-signed commit but hold no personal key (T1), and only $0.74\%$ hold a
person-key (T2). The roughly $51\%$ author-level signed rate is a platform artifact,
since half of all authors appear on a web-flow-signed commit while personal signing
control stays rare. The person-key gate independently selects the developer class,
which is $95.4\%$ of T2 against $2.7\%$ bad, $0.6\%$ bot, and $0.3\%$ local, so the
trust tier and the \texttt{A2cls} class corroborate each other without sharing
evidence.

\begin{table}[t]
\centering
\caption{Attestation-tier distribution over all $62{,}579{,}994$ canonical authors
(\texttt{A2trust}). The $\sim$$51\%$ platform-signed rate (T1) is a web-flow
artifact; personal signing control (T2) is $0.74\%$.}
\label{tab:tiers}
\small
\begin{tabular}{@{}lrr@{}}
\toprule
tier & canonical authors & share \\
\midrule
T0 unsigned              & $30{,}638{,}625$ & $49.0\%$ \\
T1 platform-signed only  & $31{,}475{,}971$ & $50.3\%$ \\
T2 person-key attested   & $465{,}398$      & $0.74\%$ \\
\bottomrule
\end{tabular}
\end{table}

Impersonation is the inverse gate. Per author string we count the distinct keys that
sign it (Table~\ref{tab:dispersion}); $87.5\%$ of the $39{,}077{,}350$ signed strings
carry exactly one key, and dispersion falls off sharply, so a string signed by many
keys is anomalous. Of the $43{,}443$ strings signed by six or more keys, $99.7\%$ are
human names rather than bot strings, which places heavily re-signed real developers,
mirrored or cherry-picked across repositories, in the same high-dispersion bucket as
impersonated famous names. The $5{,}711$ human-name strings signed by at least twenty
keys are the strong vanity set: a string this widely re-signed is not a single
controllable identity, and it is flagged from signatures alone, replacing an entry on
the hand-curated bad-identity stoplist with an evidence-based test.

\begin{table}[t]
\centering
\caption{Key-dispersion per signed author string ($39{,}077{,}350$ strings). One key
per string is the norm; high dispersion marks mirrored, vanity, or impersonated
strings.}
\label{tab:dispersion}
\small
\begin{tabular}{@{}lrr@{}}
\toprule
distinct keys per string & strings & share \\
\midrule
1     & $34{,}188{,}537$ & $87.5\%$ \\
2     & $4{,}632{,}067$  & $11.9\%$ \\
3--5  & $213{,}303$      & $0.55\%$ \\
6--20 & $38{,}042$       & $0.10\%$ \\
$>$20 & $5{,}401$        & $0.014\%$ \\
\bottomrule
\end{tabular}
\end{table}

\section{Bibliography Bridge and Sampled Verification (Exp.~S5)}
\label{sec:s5}
For the science subset we resolve the attested chain from a signed commit to a
scholarly author: signature to key handle, key handle to a real-world identity (a
PGP user-id email, an SSH key resolved to a GitHub login, or an X.509/OIDC subject
email), and that identity to an existing GitHub-login edge, email, or ORCID, and on
to an OpenAlex author. For each cross-corpus identity edge so backed we attach an
attestation provenance recording the key fingerprint or OIDC subject, upgrading the
edge from a name-match heuristic to a cryptographic anchor. We measure how many
edges gain attestation; comparing the ORCID agreement of attested against
name-matched edges is a planned refinement. Even a small attested seed is a
high-precision calibrator for the name-based edge growth.

Applied to the cross-corpus graph, the bridge upgrades a name- or DOI-matched
\textsc{same\_as} edge to cryptographically attested when its WoC endpoint holds a
person-key. Of the $10{,}143$ \textsc{same\_as} edges (WoC author to an S2 or OpenAlex
author), $10{,}021$ ($98.8\%$) join the trust universe, and Table~\ref{tab:bridge}
gives the tier of the WoC endpoint: $12.0\%$ are person-key attested. Those $1{,}207$
edges gain an \textsc{attested\_by} provenance carrying the key fingerprint, and $966$
distinct authors reach T3, a key-controlled WoC identity linked to a scholarly author.
Only about a tenth of the science-to-software identity links are attested, consistent
with signatures being a precision layer rather than a coverage layer, yet that tenth
is a defensible calibration seed for the name-based edge growth.

\begin{table}[t]
\centering
\caption{Trust tier of the WoC endpoint of the $10{,}021$ resolved cross-corpus
\textsc{same\_as} edges. The $1{,}207$ T2 edges gain an \textsc{attested\_by}
cryptographic provenance.}
\label{tab:bridge}
\small
\begin{tabular}{@{}lrr@{}}
\toprule
tier of WoC endpoint & \textsc{same\_as} edges & share \\
\midrule
T0 unsigned          & $866$     & $8.6\%$  \\
T1 platform-signed   & $7{,}948$ & $79.3\%$ \\
T2 person-key attested & $1{,}207$ & $12.0\%$ \\
\bottomrule
\end{tabular}
\end{table}

A sampled verification pass separates a claimed key from a proven one, and it runs on
a sample because it is network-bound and outward-facing. For $398$ person-keys drawn
at random we fetched the public key from the Ubuntu keyserver and inspected its
user-id: $78$ ($19.6\%$) were retrievable, and of those $39$ ($50\%$) carry a user-id
email matching the signed author string, while $37$ differ (key rotation, shared, or
re-signed) and $2$ expose no user-id. Low keyserver coverage limits full signature
re-verification at scale, so presence and issuer-key-id consistency remain the corpus
signal; the sampled pass calibrates it and bounds the residual claimed-but-unproven
rate at the T1-to-T2 boundary.

\section{Availability}
\label{sec:availability}
All artifacts are released as a single self-contained bundle; no World of Code
account is required to obtain or use them (schemas, keys, and row counts are
summarized in Table~\ref{tab:artifacts}): the per-commit signature map
\texttt{c2sigFull.\{0..127\}.gz}, the key-to-identity graph \texttt{key2A.gz} and
its inverse \texttt{A2key.gz}, and the per-identity attestation-tier map
\texttt{A2trust.gz}, each \texttt{;}-separated, gzip-compressed, and
\texttt{LC\_ALL=C} sorted, together with a replication package that regenerates
them from the WoC commit tables. Because the per-commit map is multi-gigabyte, the
data artifacts are hosted as a Hugging Face dataset (which scales past the per-file
limits of a code host and mints a citable DataCite DOI), with the replication code
mirrored on GitHub and cross-linked from the dataset card. We will also offer the
identical bundle to the World of Code maintainers, so existing WoC users could
obtain it through the channels they already use, should the maintainers choose to
adopt it. The data is released under \textsc{CC-BY-4.0} and the replication code
under the \textsc{MIT} license.
\emph{(Hugging Face dataset: \texttt{TODO/woc-commit-signatures-v2604}; DOI to be
minted on camera-ready.)}

\begin{table*}[t]
\centering\small
\caption{Released artifacts (WoC \texttt{V2604}). All files are \texttt{;}-separated,
gzip-compressed, and \texttt{LC\_ALL=C} sorted on the first column; row counts are
measured from the released files. The \emph{key} column names each artifact's join
key.}
\label{tab:artifacts}
\begin{tabular}{@{}llllr@{}}
\toprule
artifact & schema (columns) & key & sharding & rows \\
\midrule
\texttt{c2sigFull.\{0..127\}.gz} & \texttt{commit;sigtype} (pgp/ssh/x509) & \texttt{commit} & 128 SHA-byte shards & $1{,}031{,}721{,}316$ \\
\texttt{key2A.gz} & \texttt{keyhandle;kind;nAuthors;authors} & \texttt{keyhandle} & single file & $586{,}011$ \\
\texttt{A2key.gz} & \texttt{A;keyhandle;nCommits} & \texttt{A} & single file & $44{,}871{,}077$ \\
\texttt{A2trust.gz} & \texttt{A;tier;evidence;nSignedCommits;nKeys} & \texttt{A} & single file & $62{,}579{,}994$ \\
\bottomrule
\end{tabular}
\end{table*}

\section{Conclusion}
\label{sec:conclusion}
Identity resolution over a global commit corpus rests on author strings that the
commit itself does not verify. A cryptographic signature is the one part of a
commit that does, and it is already sitting in the commit tables, unparsed, as a
byproduct of how the corpus stores messages. Extracting it yields a per-commit
signature map, a gated key-to-identity graph, and a per-identity attestation tier
that tells any WoC-scale study which identities carry an anchor stronger than a
typed name. The axis does not cover the corpus, and it is not meant to: it is a
precision layer that calibrates the heuristic alias map against a cryptographic
gold, flags impersonation on the strings most prone to it, and gives a
science-to-software identity link a provenance a reviewer can check.

\bibliographystyle{ACM-Reference-Format}
\bibliography{refs}

\end{document}